\documentclass[aps,prd,preprint,tightenlines]{revtex4}
\usepackage{graphicx}

\begin{document}


\title{Updated Results from the RICE Experiment and Future Prospects for Ultra-High Energy Neutrino Detection at the South Pole}

\begin{abstract}
The RICE experiment seeks observation of ultra-high energy
(``UHE''; $E_\nu>10^{17}$ eV)
neutrinos interacting in Antarctic ice, by measurement of
the radio frequency (RF) Cherenkov radiation resulting from
the collision of a neutrino with an ice molecule.
RICE was initiated in 1999 as a first-generation prototype for
an eventual, large-scale in-ice UHE neutrino detector. 
Herein, we present updated limits on the diffuse UHE neutrino flux, 
based on
twelve years of data taken between 1999 and 2010.
We find no convincing neutrino candidates,
resulting in 95\% confidence-level
model-dependent limits on the
flux $E_\nu^2d\phi/dE_\nu<0.5\times 10^{-6}~{\rm GeV}/({\rm cm^2s-sr})$
in the energy range $10^{17}< E_\nu< 10^{20}$ eV, or approximately a 
two-fold improvement over our previously published results.
Recently, the focus of RICE science has shifted to studies of
radio frequency ice
properties as the RICE experimental hardware has been absorbed into 
a new experimental initiative (the Askaryan Radio Array, or `ARA') at South
Pole. ARA seeks to improve on the RICE sensitivity by approximately two orders of magnitude by 2017 and thereby establish the cosmogenic neutrino flux. 
As detailed herein, RICE studies of Antarctic ice demonstrate that both birefringence and internal layer RF scattering result in no significant loss of ARA neutrino sensitivity, and, for the first time, verify {\it in situ} the decrease in attenuation length with depth into the Antarctic ice sheet.
\end{abstract}

\author{I. Kravchenko}
\affiliation{University of Nebraska Dept. of Physics and Astronomy, Lincoln, NE, 68588-0299}
\author{S. Hussain, D. Seckel}
\affiliation{Department of Physics and Astronomy and Bartol Research Institute, U. of Delaware, Newark DE 19716}
\author{D. Besson, E. Fensholt, J. Ralston, J. Taylor}
\affiliation{University of Kansas Dept. of Physics and Astronomy, Lawrence KS 66045-2151}
\author{K. Ratzlaff, R. Young}
\affiliation{University of Kansas Instrumentation Design Laboratory, Lawrence KS 66045-2151}



\maketitle

\newpage

\section{Radiowave Detection of Neutrinos}
The RICE array is designed to detect events in which a neutrino-nucleon 
scattering event initiates a compact electromagnetic cascade in ice. 
Such cascades
carry a net negative 
electric charge of magnitude 0.25$e$/GeV, resulting in a pulse of radio Cherenkov emission via the ``Askaryan effect''\cite{Askaryan}, with
power peaked at wavelengths
comparable to the lateral dimensions of the cascade, i.e., the
Moliere radius ($\sim$10 cm). This paper reports new limits from the RICE array, updating the previous limits from 2006\cite{rice06}, and also discusses how recent radio frequency (RF) studies of ice properties will impact new initiatives at South Pole.

Development of radio frequency detectors to measure ultra-high energy cosmic ray interactions has recently intensified. Projects use several methods and target materials including salt\cite{SALSA}, ice\cite{RAMAND,ARIANNA,ANITAinstr,FORTE,RITA} and lunar regolith\cite{GLUE,LUNASKA}. Complementary efforts seek to measure the RF signals in cosmic ray-induced extensive air showers\cite{LOPES,CODALEMA,AERA,ANITAcr}. While calculations of RF signals from cosmic rays first appeared nearly 70 years ago\cite{BlackwellLovell_1941}, the new technology of nanosecond-scale digitizers and massive multi-channel data analysis are now bringing the potential of radio detection to fruition.

\subsection{Signal Strength}
Radio-wavelength detection of electromagnetic showers in
ice relies on two experimentally established phenomena - 
long attenuation lengths exceeding 1 km, and 
 coherence  
extending up to 1 GHz for Cherenkov emission.
Discussions of the Askaryan effect\cite{Askaryan} upon
which the radiowave detection technique is founded, its experimental
verification in a testbeam environment\cite{slac-testbeam,slac-salt04}, 
calculations
of the expected radio-frequency signal from a purely electromagnetic
shower\cite{ZHS,Alvarez-papers,SoebPRD,addendum,RomanJohn,AndresJaime},
as well as hadronic showers\cite{Shahid-hadronic}, and modifications
due to the LPM effect\cite{LPM,spencer04} can be found in the literature.

All estimates
give the same qualitative conclusion - at large distances, the signal 
at the antenna inputs
is a symmetric pulse, approximately 1-2 ns wide in the time domain. On the
Cherenkov cone, the
power spectrum rises monotonically with frequency,
as expected in the coherent 
long-wavelength limit. In that limit, the
excess negative charge in the shower front (roughly one
electron per 4 GeV shower energy) can be treated as a single
(``coherent'') source charge.
For perfect signal transmission (no cable signal losses)
through an electrically matched system, calculations estimate
the Cherenkov-cone signal strength
due to a 1 PeV neutrino initiating 
a shower at R=1 km from an antenna to be $\sim 10 \mu V\sqrt{\tt B}$, with
{\tt B} the system bandwidth in GHz. This is comparable to
the 300 K thermal noise over that bandwidth
in the same antenna, 
prior to amplification.

\section{The RICE Experiment}
Previous RICE publications described initial limits on the incident
neutrino flux\cite{rice03a}, calibration procedures\cite{rice03b}, 
ice properties' measurements\cite{RICEnZ,RICEfaraday,RICEbiref10}, and successor
analyses of micro-black hole production\cite{Shahid-hadronic},
gamma-ray burst production of UHE neutrinos\cite{grb06}, 
tightened limits on the diffuse neutrino flux\cite{rice06} and a search
for magnetic monopoles\cite{daniel}.
Herein we update both our diffuse neutrino search, based on the entire data sample, using algorithms which differ minimally from our preceding search, and also provide new information derived from radioglaciological studies of the ice dielectric permittivity at radio wavelengths.

\subsection{Experimental Layout}
\begin{figure}[htpb]\centerline{\includegraphics[width=9cm]{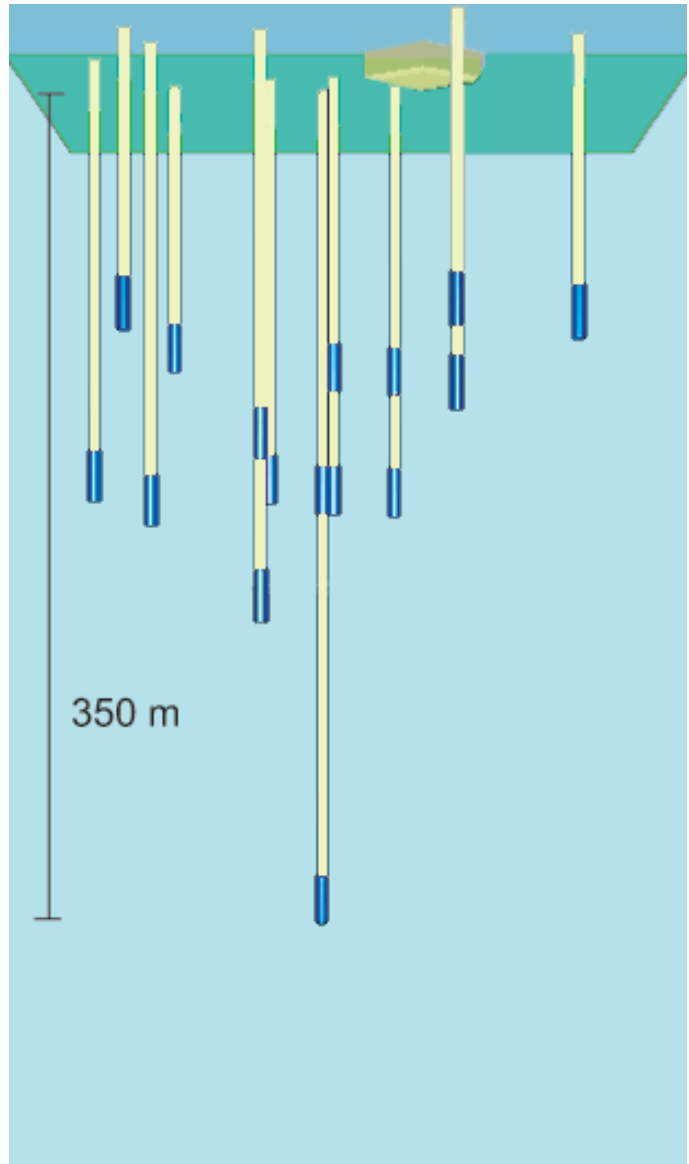}}\caption{\it Cutaway view of RICE experimental hardware. Fat-dipole antennas (show in in blue) are connected by coaxial
cables (yellow) to a data acquisition system housed in the MAPO building (shown as rectangular solid). Locations are drawn to scale relative to MAPO. As indicated in the Figure, the 
deepest antenna is approximately 350 meters below the surface.}\label{fig:viewer.eps}\end{figure}

\begin{figure}[htpb]
\centerline{\includegraphics[width=15cm]{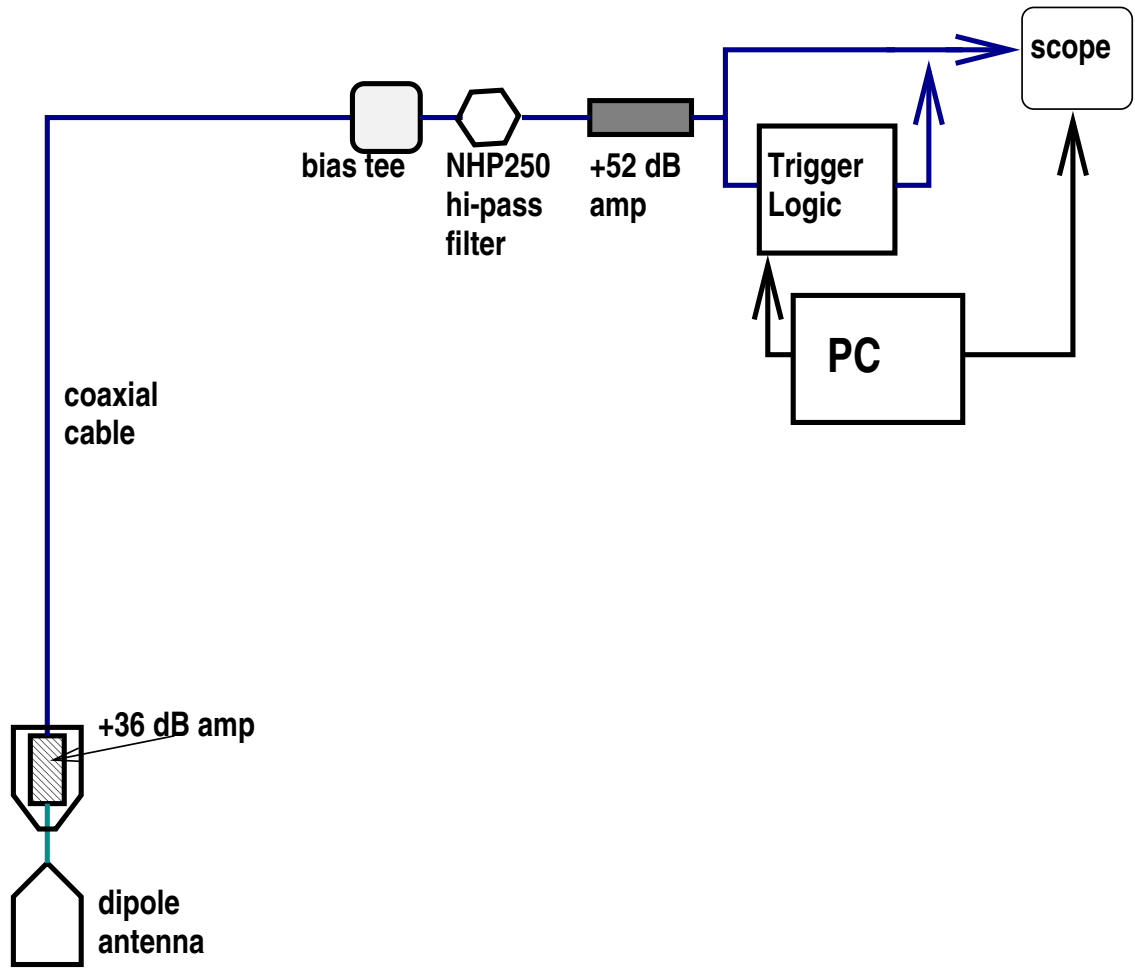}}
\caption{\it Block diagram, showing primary experimental components. Beginning with
the in-ice `fat dipole' antenna, signal is initially amplified, then conveyed by coaxial cable to the
surface, where it is high-pass filtered, and then undergoes a second stage of amplification. Signals
are then split into a `trigger' path and a `digitization' path; the latter of these brings signals into
one channel of an HP5454 digitial oscilloscope, which holds waveform data until the trigger decision is made
($\sim$1.5 microseconds). The trigger latch initiates readout of an 8.192 microsecond
waveform sample, digitized at 1 GSa/s.}
\label{fig:RICE-block-diagram.xfig.eps}
\end{figure}

\begin{figure}[htpb]\centerline{\includegraphics[width=13cm]{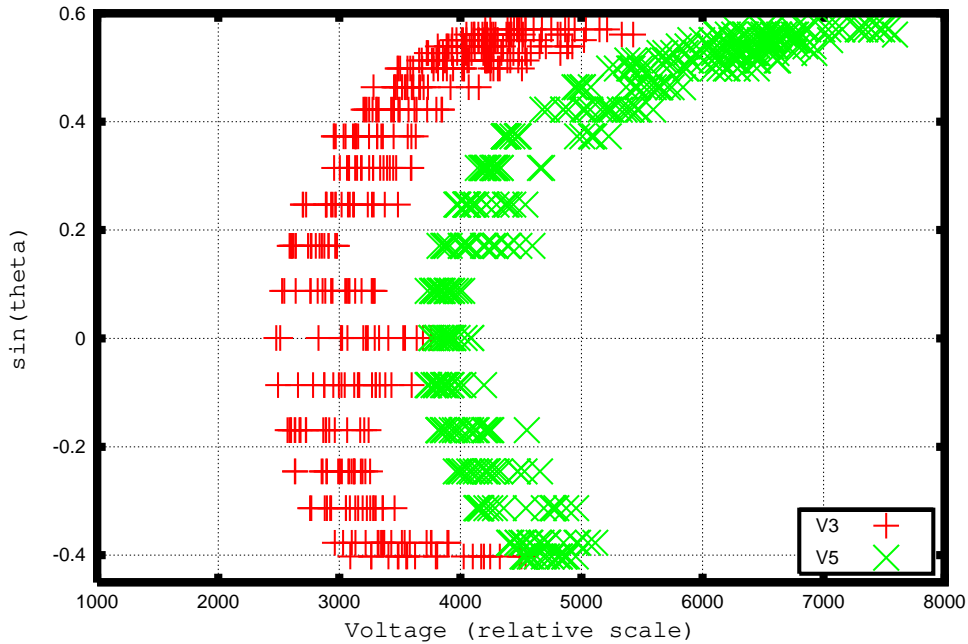}}\caption{\it Amplitude of received signal as a function of viewing angle between transmitter and receiver, for receiver channels 3 and 5 (selected on the basis of their ability to sample the largest range in $\sin\theta$). Dipole antennas, which are expected to follow a $\cos\theta$ field
beam pattern should show a reduction in a factor of 2 ($\cos^2\theta$) for broadcasts between two vertically aligned antennas, between $\sin\theta$=0 and $\sin\theta=\sqrt{2}/2$, roughly consistent with observation.}\label{fig:Rxcostheta}\end{figure}

Figure \ref{fig:viewer.eps} 
shows the detector geometry (essentially unchanged since
2000) in relation to the Martin A. Pomerantz
Observatory (MAPO) at South Pole Station.
The sensitive detector elements, 
radio receivers, are submerged at depths
of several hundred meters close to the
Geographic South Pole, in holes primarily drilled for the AMANDA experiment.
Six of the RICE receivers are deployed in `dry' holes drilled 
specifically for RICE in 1998-99. Despite bulk
motion of the ice sheet, and the closing of those dry holes
under the ambient hydrostatic pressure over $\sim$5 years,
we continue to receive signals from all successfully deployed antennas.
A block diagram of the experiment, showing the signal
path from in-ice to the surface electronics, is shown in Figure 
\ref{fig:RICE-block-diagram.xfig.eps}.
The primary sensors are the in-ice `fat-dipoles', which have good bandwidth over the frequency interval 200-1000 MHz,
and a beam pattern consistent with the expected $\cos^2\theta$ dependence for wavelengths larger than the physical scale of the dipole antenna, 
as verified in transmitter tests (Figure \ref{fig:Rxcostheta}) performed while lowering a transmitter dipole antenna into a dry borehole.

\subsection{Full Data Set \label{s:CDS}}
The statistics of the 
complete data taken thus far with the RICE array are summarized in
Table \ref{tab:datasum}. 
\begin{table}
\begin{center}
\begin{tabular}{c|cc|ccccc} \hline
Year	& RunTime & LiveTime & 4-hit Trigs & Unbiased & Vetoes \\ 
        & ($10^6$ s) & ($10^6$ s) & ($\times 10^4$) & ($\times 10^4$) & ($\times 10^4$) (prescale) \\ \hline
1999 & 0.18 & 0.10 & 0.26 & - & 1.2 (1) \\
2000 & 22.3 & 15.7 & 30.6 & 3.3 & 11182.8 (10000) \\
2001 & 4.6  & 3.3 & 6.0 & 1.3 & 317.4 (10000) \\
2002 & 19.9 & 13.6 & 16.9 & 3.5 & 12973.9 (10000) \\
2003 & 24.5 & 17.1 & 13.8 & 4.4  & 3153.9 (10000) \\
2004 & 11.6 & 9.4 & 9.4  & 2.5 & 142.5 (10000) \\
2005 & 18.3 & 15.5 & 26.5 & 4.0 & 471.0 (10000) \\
2006 & 19.3 & 16.5 & 8.9 & 4.2 & 20560.5 (10000) \\
2007 & 14.6 & 11.8 & 25.8 & 4.3 & 866.3 (10000) \\
2008 & 20.1 & 17.2 & 21.1 & 5.0 & 186.2 (10000)\\
2009 & 26.6 & 23.8 & 10.1 & 8.3 & 488.2 (10000)\\ 
2010 & 23.1 & 21.9 & 6.1 & 5.5 & 224.4 (10000) \\ 
\hline
\end{tabular}
\caption{\it Summary of RICE data taken through December, 2010.
``4-hit Triggers'' refer to all events for which there are
at least
four RICE antennas registering voltages
exceeding a pre-set discriminator threshold 
in a coincidence time comparable to the light transit time across the array ($1.25\mu$s); ``Unbiased Triggers'' correspond to the
total number of events taken at pre-specified intervals
and are intended to capture
background conditions within the array; 
``Veto Triggers'' are events tagged online by a fast ($\sim$10 ms/event)
software algorithm as consistent with 
having a surface origin. With the cessation of
AMANDA operations in March, 2009, beginning in February 2010,
the ``AMANDA'' trigger line was replaced by a 3-fold
surface-antenna multiplicity trigger. Variations in the veto rate are attributed to the commissioning of new experiments, with associated anthropogenic electromagnetic interference, and also the decommissioning of other experiments, as well as communications streams such as the GOES satellite in 2006.}
\label{tab:datasum}
\end{center}
\end{table}
Over a typical 24-hour period, roughly 1500 data
event triggers pass a fast online hardware surface-background veto (``HSV''; with a decision time $\sim$5$\mu$s/event) 
and an online software surface-background
veto ($\sim$10 ms/event). To these data we have applied a sequence 
of offline cuts to remove background, as detailed below.

\section{Trigger and Data Collection} 
Our basic online procedures are essentially unchanged from our previous
publication. The first three tiers, or trigger levels (Table \ref{tab:trigsum}) are applied in either hardware (H) or software (S)
online, as follows:
\begin{enumerate}
\item L0: Passes Hardware Surface Veto, with one antenna exceeding a threshold approximately equal to six times the ambient background noise level.
\item L1: Four antennas satisfying an L0 requirement within a coincidence
time window equal to the light transit time across the
array (1.25 microseconds).
\item L2: Events are deemed to be inconsistent with originating at the surface, using a software veto.
\end{enumerate}
Events passing all tiers are transferred daily from the South Pole
for permanent storage on disk at the University of Wisconsin.
\begin{table}[htpb]
\begin{center}
\begin{tabular}{c|c|c|c} \hline 
Trigger Level & Requirement & Maximum Rate & Typical Winter Rate \\ \hline
L0 (H) & Passes HSV veto & 200 kHz & 1 kHz \\ 
L1 (H) & Four antennas exceed threshold within 1.25 $\mu$s & 100 Hz & 1 Hz \\
L2 (S) & Passes surface-source veto & 0.1 Hz & 0.02 Hz \\ \hline
\end{tabular}
\caption{\it Summary of trigger rates at three levels. ``HSV'' refers to the online
hardware surface veto of down-coming, anthropogenic noise. The third column represents the 
maximum trigger rate to process events exceeding the online
event threshold. The final column represents the maximum rate at which data can be 
written to disk.}
\label{tab:trigsum}
\end{center}
\end{table}

\section{Hit Finding and Event Reconstruction}
\begin{figure}[htpb]
\centerline{\includegraphics[width=10cm,angle=0]{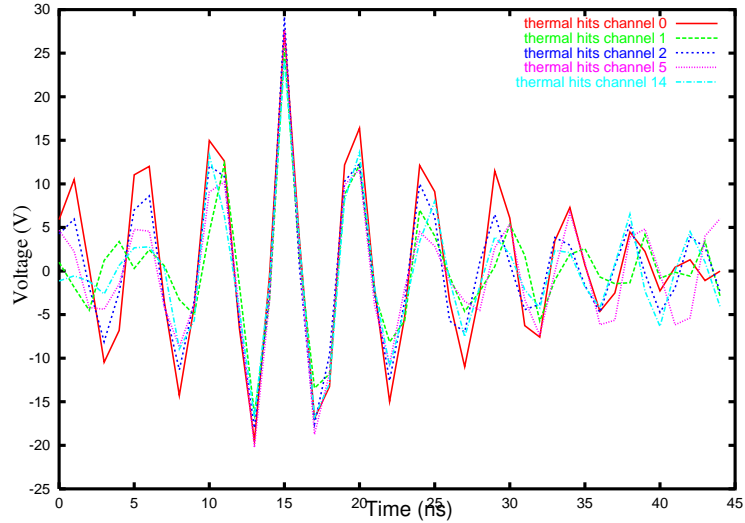}}
\caption{\it ``Short-duration'' waveforms, 
selecting
cases with fast impulsive responses (designated as
``thermal hits''), for the indicated channels.}
\label{fig:sigwvfm0}
\end{figure}

\begin{figure}[htpb]
\centerline{\includegraphics[width=10cm,angle=0]{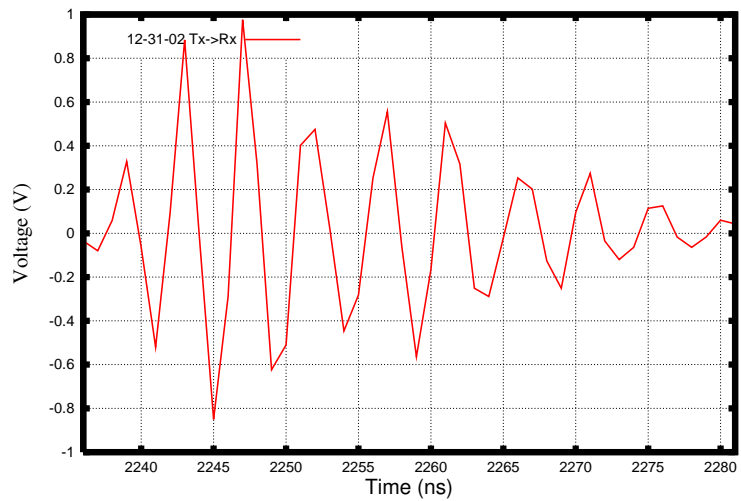}}
\vspace{0.5cm}
\caption{\it Receiver waveforms captured when transmitter is active, 
 for comparison with 'thermal' hits in previous Figure.}
\label{fig:sigwvfm6}
\end{figure}

Our previous analysis assigned the hit time to the first
$6\times\sigma_V$ excursion in a waveform, with $\sigma_V$ defined as the rms of the
voltages recorded in the first 1000 ns of an event capture (i.e., prior to
the possible onset of any signals).
To improve our hit-finding,
we have performed a study of time domain signal characteristics. 
An ideal narrow-band antenna has a response to an impulse
that follows the form:
$$V_0(t)\sim\cos(\omega_0t)exp(-t/\tau),$$ 
with $\tau\sim$1/{\tt B}. 
For a first approximation
to what these signals might look like, we used `thermal noise hits', defined
as a sequence of waveform samples in unbiased events consistent with band-limited transients, drawn
from the data itself.
Figure \ref{fig:sigwvfm0} 
qualitatively indicates the reproducibility of such short-duration transient
responses from channel-to-channel (2002 data),
with the response to a sharp transmitter signal shown in Figure
\ref{fig:sigwvfm6} for comparison. 

We have performed an embedding study to evaluate contributions to hit time
resolutions and the relative efficacy of the damped exponential parametrization
to previous hit time
definitions. Monte Carlo simulations of neutrino-induced hits are
embedded into data unbiased events, and the extracted hit times then
compared with the known (true) embedded time.
\begin{figure}[htpb]\centerline{\includegraphics[width=10cm,angle=0]{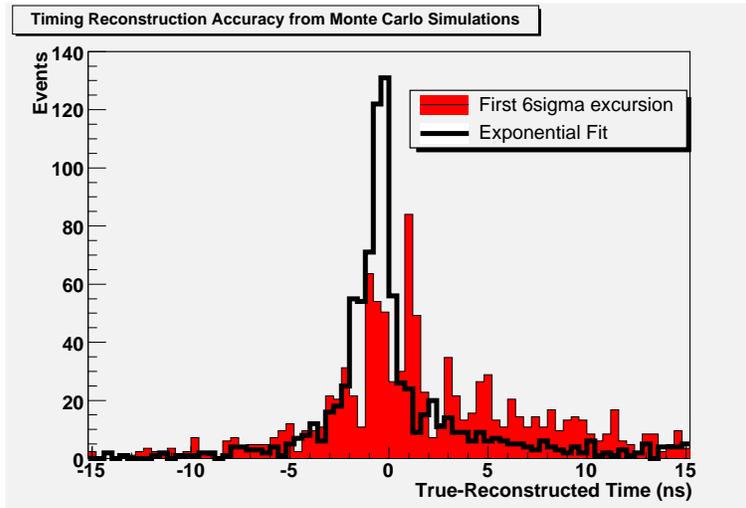}}
\caption{\it Difference between the true (embedded) time between antenna hits minus the reconstructed hit time (after embedding) for two reconstruction algorithms. The indicated timing resolution due to pattern recognition uncertainties is approximately 2 ns.}
\label{fig:dt-embedding}\end{figure}
Our previous analysis employed 
the `first 6$\sigma_V$ excursion' hit criterion; based on Figure \ref{fig:dt-embedding} and the signals shapes displayed previously, we have now applied the more general `exponential ring' hit criterion, which records the time of waveforms exceeding 
$5.5\sigma_V$ 
in voltage, and also have the shape of a damped exponential `ring'. Monte Carlo simulations
indicate that this improved timing resolution
improves the azimuthal directional resolution of RICE by approximately 10\%.

\begin{figure}[htpb]\centerline{\includegraphics[width=10cm]{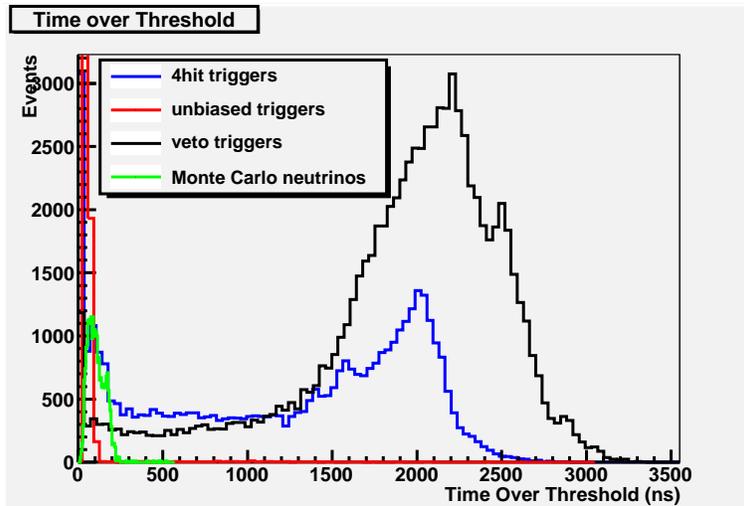}}\caption{\it ``Time-over-threshold'' for 2010 event triggers, by trigger type, as described in the text.}\label{fig:ToTToT.eps}\end{figure}

Such an algorithm is also designed to reject events where the signal persists for hundreds of nanoseconds in each
channel. Figure \ref{fig:ToTToT.eps} shows the time-over-threshold distribution, defined relative to the
rms voltage $\sigma_V$, as the number of samples
exceeding $\pm6V_{rms}$, 
for various triggers. The contamination of our `general
trigger' data sample with large time-over-threshold
events, which nearly uniformly trace to the surface, is evident from Figure \ref{fig:ToTToT.eps}. 
Such waveforms are immediately
rejected by requiring the expected damped exponential signal form.
Also overlaid (green) is the distribution expected from Monte Carlo simulations, exhibiting a considerably narrower distribution and clustered towards zero time-over-threshold.

\section{Backgrounds\label{s:bkgnds}}
Our previous publication\cite{rice06} presented detailed consideration of
anthropogenic transients, thermal noise backgrounds, 
and possible backgrounds from atmospheric muons, atmospheric
neutrinos, air showers, and RF emissions due to solar flares.
We herein briefly review techniques for background suppression.

We generally distinguish different backgrounds to
the neutrino search according to the following
criteria:
\begin{itemize}
\item vertex location of reconstructed
source \item waveform shape characteristics of
hit channels (including, e.g., time-over-threshold [Fig. \ref{fig:ToTToT.eps}])
\item goodness-of-fit to a 
well-constrained single vertex as evidenced by
timing residual characteristics (defined as the inconsistency of the
recorded hit channels to originate from a single source point)
\item RF conditions during data-taking
\item Fourier spectrum of hit channels 
\item cleanliness of hits
(e.g., presence of multiple
 pulses in an 8.192 microsecond waveform capture)
\item multiplicity of receiver antennas registering hits for a particular event
\item time-since-last-trigger 
($\delta t_{ij}\equiv t_i-t_j$, where $t_i$ is the time of the i$^{th}$ trigger and $t_j$ is the time of the next trigger). In high-background, low-livetime instances, we expect  $\delta t_{ij}\to \delta t_{min}$, where $\delta t_{min}$ is the $\sim$10 s/event readout time of the DAQ. In low-background, high-livetime instances, we expect $\delta t_{ij}\to \delta t_{max}$, where $\delta t_{max}$ is the ten-minute interval between successive unbiased triggers.
\end{itemize}
We can coarsely
characterize three general 
classes of backgrounds according to the
above scheme, as follows. 

1) Continuous wave backgrounds (CW) have
a) a long time-over-threshold
for channels with amplitudes well above the
discriminator threshold, b) large timing residuals (since the threshold
crossing times will
be ambiguous), c) small values of
$\delta t_{ij}$ for the case where the discriminator threshold is
far below the CW amplitude, 
d) a Fourier spectrum dominated by one frequency (plus overtones),
e) a hit multiplicity which is on 
average roughly
constant, and determined by the number of channels which
exceed threshold when their noise voltage is added to the underlying
CW voltage. Such backgrounds may cluster in time 
and are generally easily recognized on-line.

\begin{figure}[htpb]\centerline{\includegraphics[width=10cm]{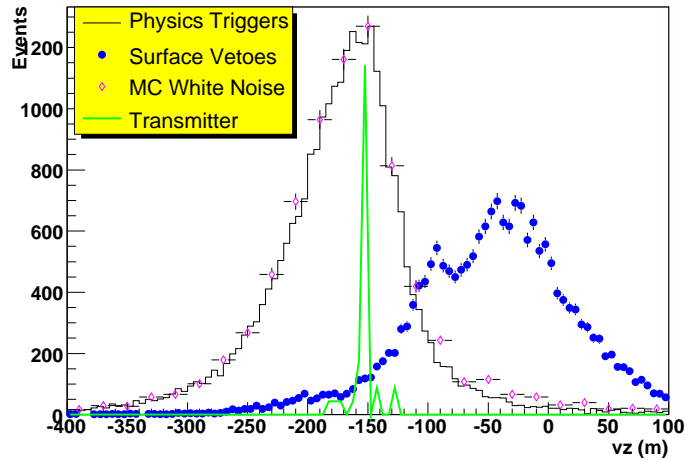}}\caption{\it Reconstructed source depth for primary neutrino
search triggers (``Physics triggers'') compared to 
events identified as surface sources based on Hardware Surface
Veto information, as well as transmitter calibration events (green) and 
simulated thermal noise (points).
When anthropogenic backgrounds are low
and the experiment is operating close to the thermal limit, the
reconstructed vertex distribution for thermal noise
events is expected to peak close to the
center of the array, with a width given by the light transit distance
across the 1.25 $\mu$s
coincidence window defined by the RICE general event trigger.
During the winter months, when station noise is typically lowest, 
approximately 50\% of recorded events are thermal noise backgrounds.
During the austral summer months, when human activity at
South Pole Station is largest, this fraction typically decreases to less than
10\%.}\label{fig:plot_thermal_2000.eps}\end{figure}

2) True thermal noise backgrounds should have a) three-dimensional
vertex locations which are spatially distributed as Gaussians  
peaked at the centroid of the array ($x$=0, $y$=0, $z$=--120 m), as demonstrated by
Monte Carlo simulations (by simulating four hits at random times 
within the 1.25$\mu$s discriminator window;
see Fig. \ref{fig:plot_thermal_2000.eps}),
b) very small time-over-thresholds, c) large timing
residuals, d) successive
trigger time difference characteristics which depend in a statistically
predictable way on the ratio of discriminator thresholds to rms 
background noise
voltages, 
e) a ratio of general/unbiased triggers which, in principle, can be
statistically derived from the background noise distribution observed in 
unbiased events, f) a Fourier spectrum determined 
by the intrinsic band response
of the various components of a RICE receiver circuit, 
g) no double pulse characteristics, 
h) no correlation with date or time. 

3) ``Loud''
transients are observed to constitute the
dominant background.
We sub-divide possible transient
sources into two categories: those sources which originate within the ice
itself, primarily due to AMANDA and/or
IceCube photomultiplier
tube electronics, and those
sources which originate on, or
above the surface. 
After our initial deployment of three
test antennas in 1996-97, highpass 
($>$250 MHz) filters were inserted
to suppress the former backgrounds, leaving 
more sporadic anthropogenic
surface-generated noise as the dominant transient background.
Such triggers
are characterized by: 
a) typically, large time-over-thresholds, 
b) $\delta t_{ij}$ distributions which reflect saturation
of the DAQ data throughput, or 
show time structure if the source is periodic, c) Fourier spectra which 
are likely to depart from thermal ``white'' noise in the frequency domain.

\subsection{Vertex
Suppression of Transient Anthropogenic Backgrounds \label{s:TAB}}

Vertex distributions
give perhaps the most direct characterization 
of surface-generated (z$\sim$0) vs.
sub-surface (and therefore, candidates for more interesting processes) events.
Consistency between various source reconstruction algorithms gives 
confidence that the true source
has been located.
Due to ray tracing
effects, it is difficult to identify surface sources at large
polar angles, which increasingly fold into the region around
the critical angle.
We implement both a ``grid''-based $\chi^2$
vertex search algorithm, as well as an
analytic, 4-hit vertex reconstruction algorithm, as 
detailed previously\cite{rice03a}.
We have additionally cross-checked our vertex-finding against
results obtained using the CERN Minuit package. 

In our offline analysis, we require that the
reconstructed vertex depth be greater than 200 meters to
suppress anthropogenic surface noise.
 
\subsection{Vertex Quality Requirements}
We impose a maximum time residual (defined as the time
deviation from consistency of
the recorded antenna hit times with a single in-ice source
point) requirement of less than 50 ns, per antenna hit. Since four antennas will
necessarily allow a solution of the equation ${\bf r}=c(t-t_0)$, with
$t_0$ the time of source emission,
imposition of this requirement necessitates a minimum antenna hit
multiplicity of five. This requirement is particularly effective at
removing events where there may be a thermal noise fluctuation superimposed on
anthropogenic noise, or multiple anthropogenic events which overlay upon
each other.

\subsection{Rejection of repetitive patterns}
Inconsistency of the recorded hit time sequence with a previously
logged hit time sequence, irrespective of the previous two requirements can
also be used to identify backgrounds.
This final requirement requires one full pass of each year's data to
create a `library' of identified
background hit time patterns, as follows. As each
new event is processed, 
if the sequence of hit antennas for that event matches a previously
recorded pattern to within 10 ns per antenna,
then: a) the event is considered to be
`repetitive' and is discarded from further signal candidacy, and b) the
pattern itself is updated with a statistical weighting of the new
event with all the previous events identified as consistent with that
pattern. As a concrete example, an event with hit times in the first
four antennas (exclusively) of 100, 250, 400, and 600 ns would be
`clustered' with a previously logged pattern (with statistical
weight 1) in the exact same
channels with hit times of 92, 258, 403, and 605 ns, resulting in a
modified `clustered' pattern, with
statistical weight of 2, and re-weighted hit times 96, 254, 401.5, and 602.5 ns.

The loss in neutrino efficiency incurred by
this `template cut' algorithm is assessed by simulation to be of
order 1\%.


\subsection{Air Shower Backgrounds}
There are possible radio signals associated directly with cosmic 
ray air showers. These include the production of geo-synchrotron 
radiation in the atmosphere, as well as transition radiation 
and Cherenkov signals 
produced as the shower impacts and evolves into the ice. These three 
mechanisms all require coherent radiation from all or part of the shower. 
In all three cases, the transverse profile of the shower dictates a 
fundamental frequency response, whereas for the geo-synchrotron and 
Cherenkov signals the shower/observer geometry must also be favorable to 
have coherent emission from the full longitudinal development of the 
shower.

Coherent production of synchrotron radiation in the geomagnetic field 
has recently been observed by the LOPES\cite{LOPES}, 
CODALEMA\cite{CODALEMA}, AERA\cite{AERA},
and ANITA\cite{ANITAcr} 
collaborations. The coherent air shower signal is most intense
below 100 MHz\cite{HuegeRefs}, but, as demonstrated by ANITA, still detectable in the RICE bandpass,
which may attest to the observability of the air Cherenkov pulse that accompanies the geosynchrotron signal.
We have not studied this mechanism in detail, but note that the frequency 
response is ultimately related to the geometry of the air shower -- the 
signal rolls over at $f\sim cR/r_M^2$ where R$\sim$1 km is the 
height of shower max and $r_M \sim$100--200~m is the Moliere radius for the 
shower.

Transition radiation results when the shower impacts the ice\cite{gazazian}. 
In this 
case, R$\sim$200~m for RICE, f$\sim$200~MHz, and the region for 
coherent emission is a disk of order 10~m radius. Only a fraction of the 
excess shower charge, typically 10\%, is contained within that distance of the shower axis. 
Further, transition radiation is forward peaked, so illumination of more 
than one antenna string is rather unlikely. We have not seriously modeled  
transition radiation from air shower impacts as a background for RICE.

The most interesting signal for RICE is the Askaryan pulse produced when 
the air shower core hits the ice. At RICE frequencies, the Askaryan pulse 
must originate from a transverse dimension comparable to that for a 
shower initiated in-ice, a few tens of cm at most. This length scale is 
compatible with the core of the shower where the highest energy particles 
reside. Particles have their last interactions of order 1 km above the 
ice, so the required relativistic-$\gamma$ factor is of order $10^4$, 
corresponding to surface
particle energies $\sim 10$~GeV for $e^-$, $e^+$ and 
bremstrahlung $\gamma$'s. 

\begin{figure}[htpb]\centerline{\includegraphics[width=14cm]{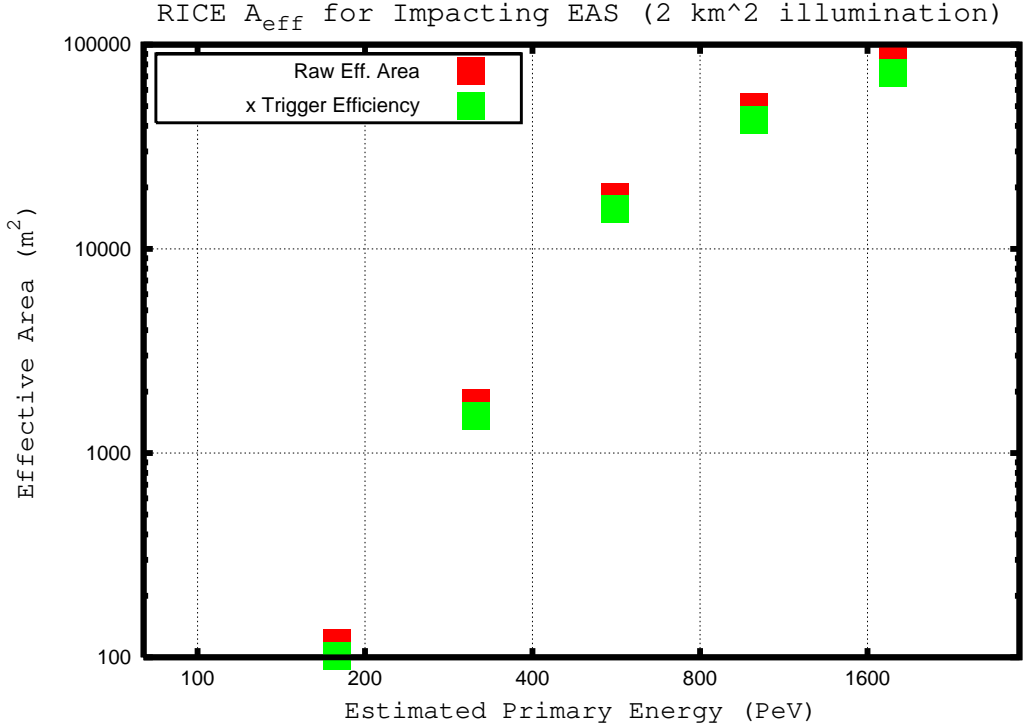}}\caption{\it Monte Carlo simulation results for RICE effective area for air showers impacting the South Polar surface.}\label{fig:EffAreaEAS.eps}\end{figure}

We have run the standard RICE Monte Carlo simulation to assess the acceptance to impacting air shower cores. Simulated events illuminate the surface isotropically from the upper hemisphere over a distance within 500 meters of the center of the RICE array. At large zenith angles, the likelihood of
four antennas being within some portion of the Cherenkov cone becomes large, however, the practical
ability to separate such signals from surface
background near the horizon is diminishingly small.
Figure \ref{fig:EffAreaEAS.eps} displays the corresponding effective area, as a function of shower core energy,
assuming the 10\% shower core containment cited above.
Given that the charged cosmic ray flux is approximately 1/5000~$m^2$/yr at the nominal RICE
event detection threshold of 100 PeV, and falling with an $E^{-2.7}$ power law (so that the integral flux falls as $E^{-1.7}$), Figure \ref{fig:EffAreaEAS.eps} indicates that the expected detection rate per year for RICE is likely to be undetectably small.


Such a possible signal was, in fact, explicitly rejected
as consistent with down-coming anthropogenic noise, prior to 2009. In 2009,
that veto was somewhat loosened, specifically to admit such possible signal.
For this search,
we required a `direct hit' corresponding to a vertex within 20 m of the center of the RICE array
in order to have any opportunity at imaging the down-coming Cherenkov ring itself and thereby 
unambiguously discriminate against above-surface backgrounds;
such backgrounds, at large
zenith angles, will fold into a tight polar angle region around the critical angle $\theta_{crit}$.
Imposition of this fiducial requirement, of course, limits
the effective area to a maximum of $\sim$1000 $m^2$. 

For this search,
all data were processed through a separate analysis chain with minimal initial
event selection requirements consisting of: i) a reconstructed impact 
point within that allowed fiducial area, 
ii) good agreement between the two vertex finders,
and iii) a very loose ``cut'' on the goodness-of-fit to a Cherenkov cone, requiring that
the event $\chi^2$ be less than 100. These requirements allow only six event candidates in all of
the 2009 data. Unfortunately, all six events fail subsequent
time-over-threshold requirements on
the waveform shape, resulting in no down-coming impacting air shower candidates.

\subsection{Wind Backgrounds}
\begin{figure}[htpb]\centerline{\includegraphics[width=1.1\textwidth]{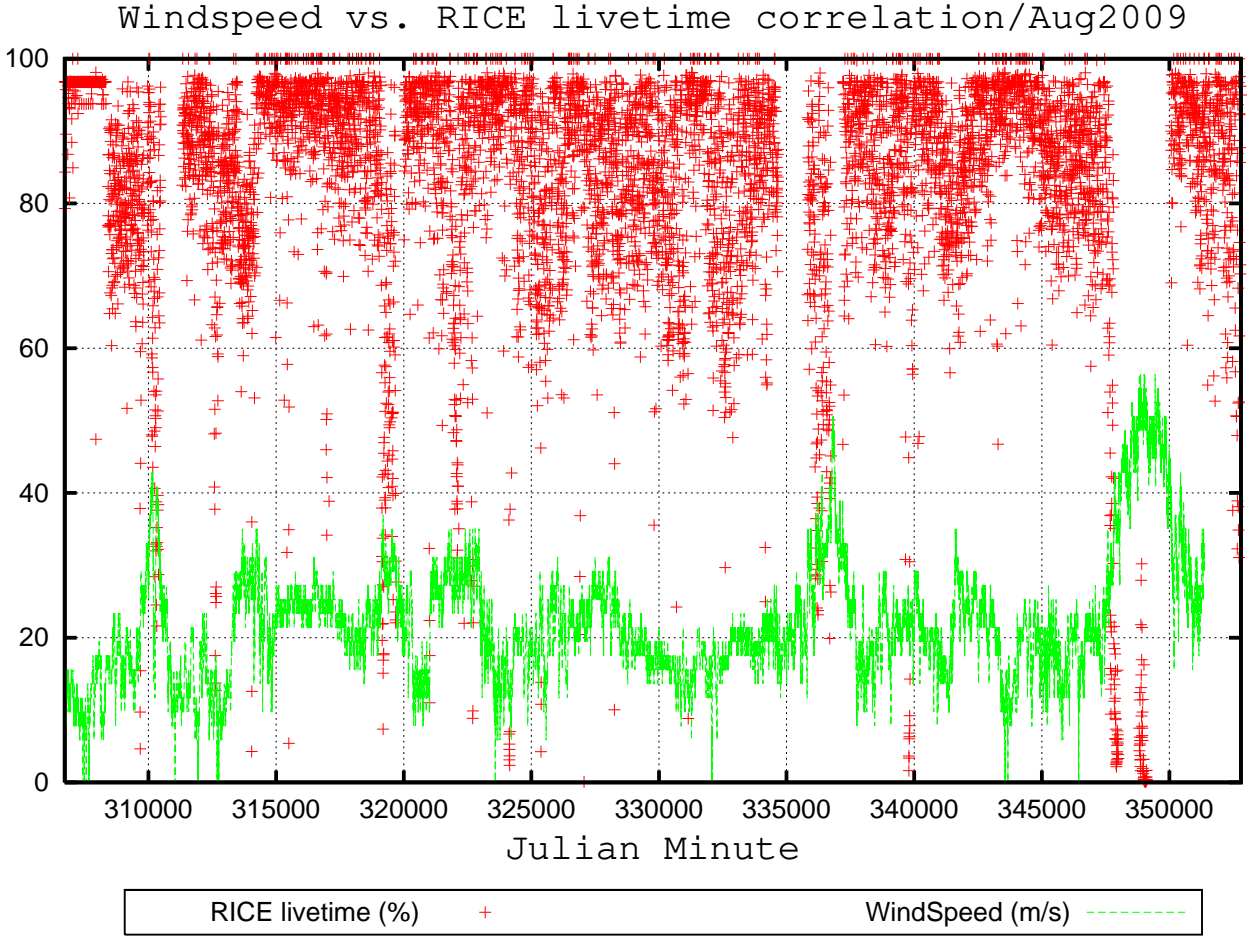}}\caption{\it Tabulated windspeed at South Pole (m/s; green) vs.
RICE livetime (red). High windspeeds apparently result in large
electrostatic discharge events from local above-surface
structures.
Such events are flagged offline as of non-neutrino origin.}\label{PlotWindSpeedCorr.eps}\end{figure}

The extraordinary lack of moisture at South Pole, coupled with a volatile
surface snow layer and an absence of large conductors to facilitate discharge
of atmospheric electrostatic fields, results in extremely high atmospheric breakdown voltages from above-surface
structures. Surface charge build-up can be enhanced by
high wind velocities. Rapid discharges can subsequently produce measurable radio frequency signal. As demonstrated in  
Fig. \ref{PlotWindSpeedCorr.eps}, we observe an apparent correlation of
trigger rate with wind velocity, as expected in a surface discharge model.
Fortunately, these events typically trace back to the surface and do not pose
an in-ice neutrino background.

\subsection{Ambient (non-episodic) Radio Frequency Backgrounds at Pole}
\begin{figure}[htpb]\centerline{\includegraphics[width=10cm]{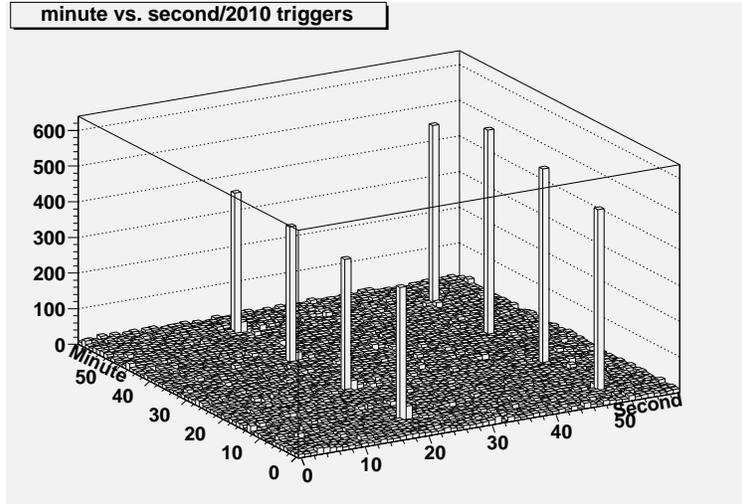}}\caption{\it Trigger time (second vs. minute) showing the 15 minute periodicity of the Very Low Frequency radar system, operating at 19.5 kHz, at South Pole.}\label{fig:VLF.eps}\end{figure}

The Very Low Frequency (VLF)\cite{VLF} receiver array at South Pole is intended to monitor the ionosphere using a large set of buried antennas, at frequencies well below the RICE sensitivity. Nevertheless, the high power of the signal broadcast by this array can evidently couple into the electronics of the RICE data acquisition system, resulting in a measurable number of triggers (10.6\% of all our physics triggers in 2010, e.g. [Figure \ref{fig:VLF.eps}]).
The waveforms in such events, however, immediately 
fail our exponential ring criterion in the offline analysis.
The VLF background is by far the most pernicious of the periodic backgrounds observed to contaminate the RICE data sample.

\section{Monte Carlo simulations}
We determine the neutrino detection efficiency of our
event selection criteria using simulations of showers, both
electromagnetic and hadronic, resulting from neutrino
collisions, superimposed on environmental characterization drawn 
from data itself at random times (unbiased events). 
Our basic Monte Carlo simulation signal codes are unchanged since 2005, save for updates to modeling radio frequency ice dielectric response (detailed below). In practice, due to the LPM effect, our
sensitivity to neutrino interactions is dominated by the response to hadronic showers, and is
therefore approximately uniform for all three neutrino flavors.

\subsection{Event-finding efficiency}
Our overall event-finding efficiency is approximately unchanged
from our prior estimate. Of simulated events which
we expect to trigger the RICE detector,
we expect 74.2\% to pass our primary event selection requirements.
Cut-by-cut details are presented in Table \ref{tab:MCeff}. Note that the definition
of the cuts generally follows our previous analysis, with only slight differences.

\begin{table}[htpb]
\begin{center}
\begin{tabular}{c|c|c} \hline
Requirement & Efficiency (\%) & Data Events \\ \hline
Starting sample & 100.0 & 2298921 \\
$\ge$4 6-sigma hits &        99.9 & 1754982 \\
Maximum time-over-threshold cut & 99.7 & 565891 \\ 
$\le$two channels with high time residual & 98.0 & 145723 \\
Acceptable total time residual  & 95.5 & 38922 \\
Passes amplitude template cut & 92.1 & 8035 \\
Passes time template cut & 89.1 & 1043 \\
Vertex of at least one algorithm below firn & 88.3 & 279 \\
Agreement between two vertex-finding algorithms & 85.1 & 140 \\
Passes Cherenkov cone geometry cut & 81.8 & 36 \\
5 high quality 6$\sigma_V$ hits & 77.4 & 8 \\
Deepest channel hit first & 77.2 & 0 \\ 
No surface antennas with good `early' hits & 74.2 & 0 \\
\hline
\end{tabular}
\caption{\it Cumulative Monte Carlo efficiency, using simulated neutrino events embedded into forced trigger events. Fractional efficiencies are
measured relative to a total of 2500 simulated events, assuming
1:1:1 mix of $\nu_e:\nu_\mu:\nu_\tau$, which passed our simulated trigger criteria.  Also shown are event survival statistics for RICE data. Note that these event selection criteria are designed to encompass, and reject, all possible backgrounds itemized in the text, 
without necessarily targeting just one type.}
\label{tab:MCeff}
\end{center}
\end{table}

\subsection{Effective Volume}
\begin{figure}[htpb]\centerline{\includegraphics[width=13cm]{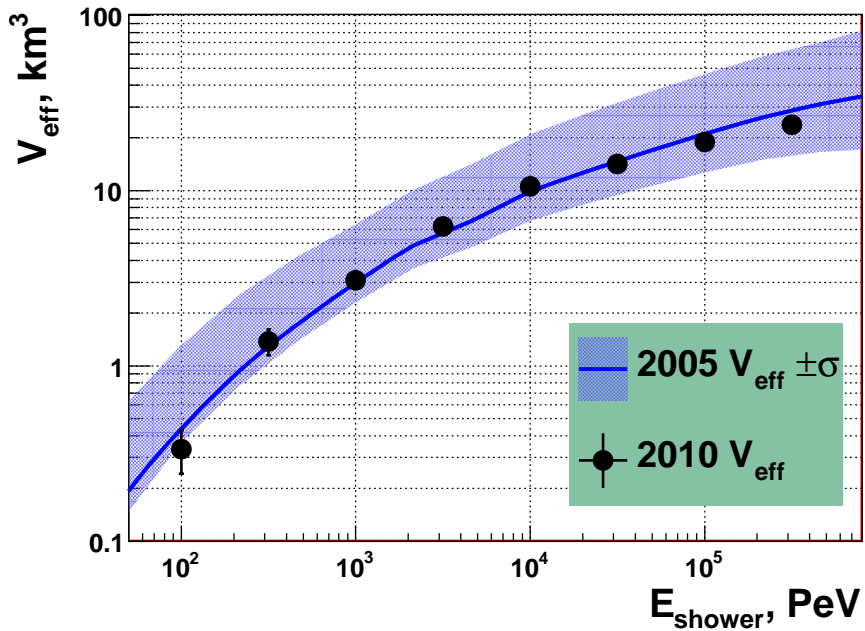}}\caption{\it Comparison of effective volume calculated with current Monte
Carlo simulations with effective volume calculated for results reported in 2005.}\label{fig:Veff0.eps}\end{figure}

\begin{figure}[htpb]\centerline{\includegraphics[width=15cm]{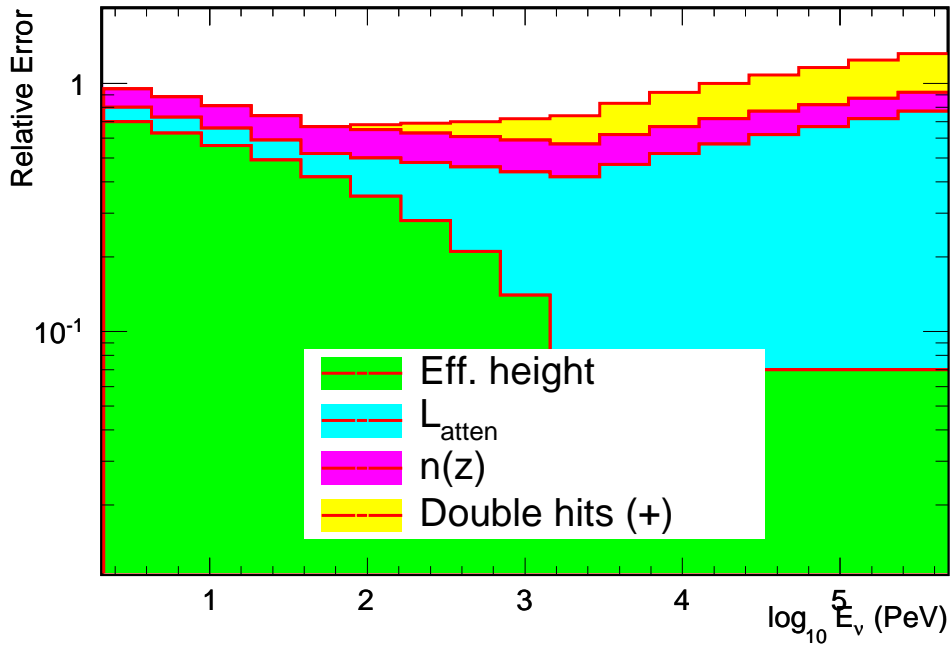}}\caption{\it Contribution to RICE systematic uncertainties in effective volume, as a function of energy, as detailed in the text.}\label{fig:SysErr}\end{figure}

As can be seen from Figure \ref{fig:Veff0.eps}, the overall neutrino response, as measured by effective volume, is
essentially unchanged relative to our previous analysis.
Systematic errors
in effective volume, as indicated by the shaded band
in Figure \ref{fig:Veff0.eps}, result in
roughly a factor of two possible variation in the 
expected overall neutrino event yield.
The breakdown of the contribution of various systematic uncertainties to
our total systematic error is very similar to our previous analysis, with the 
exception of improvements in our understanding of ice properties, as outlined elsewhere in this document.
These result in a net improvement of approximately 15\% in total systematic error compared to our previous publication. Dominant systematic errors remain uncertainties in the attenuation length as well as uncertainties in the index-of-refraction profile through the firn. Figure \ref{fig:SysErr} shows the relative contribution of various parameters to the overall systematic error. Shown are components due to
uncertainties in the effective height (green), radiofrequency attenuation length of the ice (cyan), uncertainties in the index-of-refraction (magenta, and dominant in the upper 200 m of the ice sheet), and also uncertainties due to the possiblity of Cherenkov signals generated in-ice, which reflect back down off the ice-air interface, and intercept the RICE array from above (yellow). Note that the latter effect, which can lead to so-called `double' hits, only increases the estimated effective volume, and is (conservatively) excluded from our calculation of the flux upper limit in the current analysis.

\section{Search for in-ice Neutrino interactions and Discussion}
\begin{figure}[htpb]\centerline{\includegraphics[width=10cm]{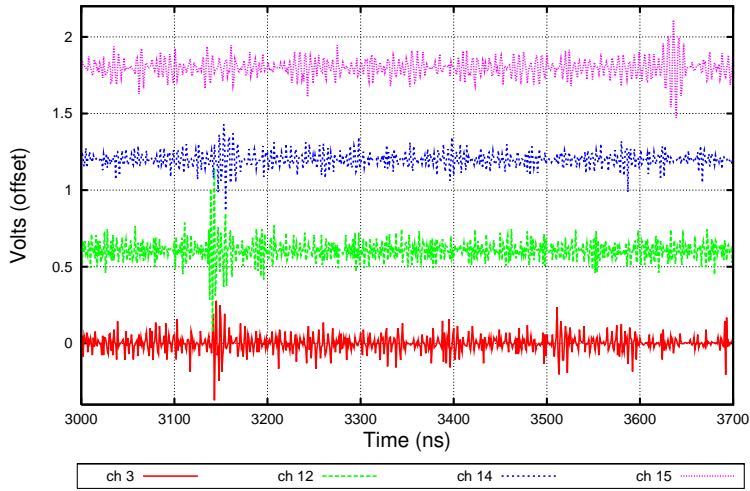}}\caption{\it Waveform display from event taken on Julian Day 93, UTC 02:45:41.407834.6 (2005). This event satisfies all event criteria listed in Table \ref{tab:MCeff}, save for the requirement that the deepest channel (Ch. 15) have a hit time preceding, rather than following the other hits in the array. As can be seen from the Figure, the late impulse observed on Channel 15 marks this event as originating from the surface.}\label{fig:disp}\end{figure}

Imposing the event selection requirements enumerated in Table \ref{tab:MCeff}, we find that
no events survive as in-ice shower candidates. 
One of the few events which satisfied all the waveform characteristic requirements, but was flagged as
having surface origin is shown in Figure \ref{fig:disp}. 

\subsection{Neutrino Flux Limit Results \label{s:NFLR}}
\begin{figure}[htpb]\centerline{\includegraphics[width=17cm]{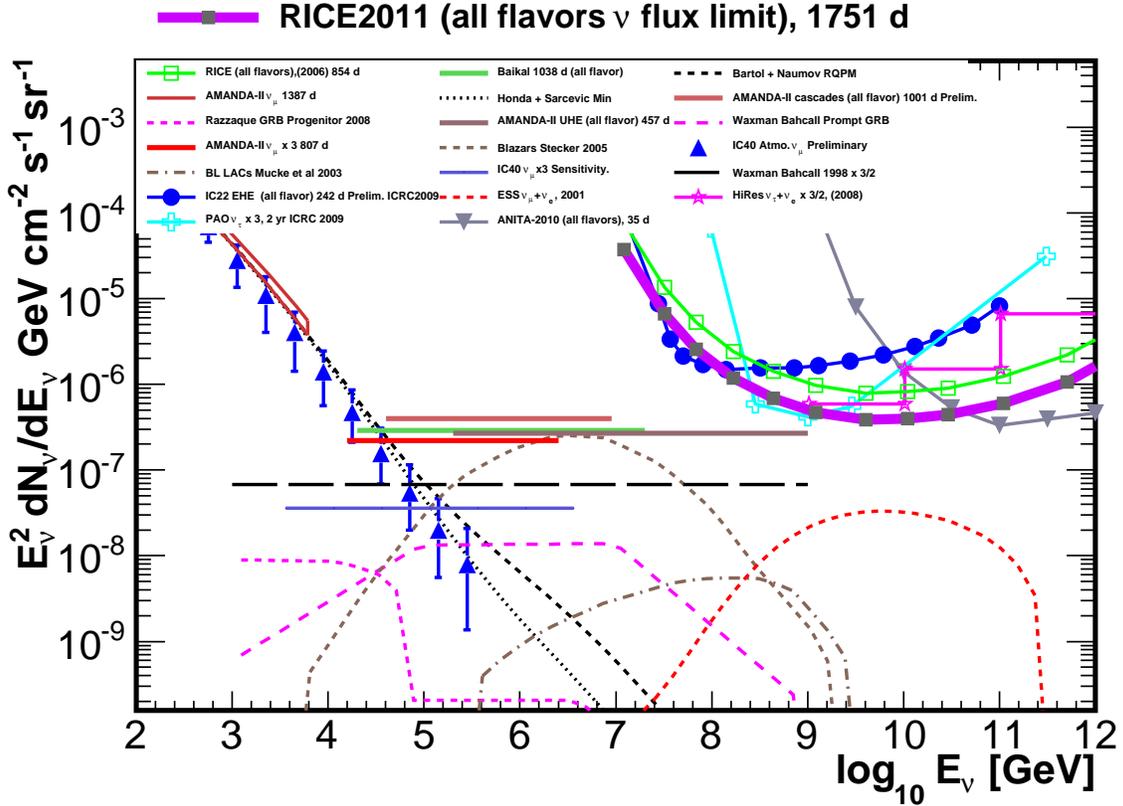}}\caption{\it 
Compilation of existing 
neutrino flux limits, including updates reported herein. 
Factors of 3 or 3/2 shown in the plot are needed to translate 
predictions and experiments sensitive to only one or two neutrino 
flavors to the three flavors of neutrinos to which the RICE experiment is sensitive.
Model predictions shown are from calculations of the neutrino fluence from
blazars by Stecker\cite{Stecker2005}, BL LAc galaxies\cite{Mucke2003},
GRB's\cite{Soeb2003}, photonuclear production of neutrinos by cosmic-ray interactions with the 
Cosmic Microwave Background\cite{ESS2001}, models of neutrinos generated locally to Earth\cite{Ina2008,Naumov2001,Bartol2004,Honda2006} and Active Galactic Nuclei fluence predictions\cite{Julia2005}. Other presented experimental limits are those from
AMANDA-II\cite{JohnKelly2009,AMII09,AMII10,AM08}, the previous RICE result\cite{rice06}, 
the HiRes experiment\cite{HiRes}, based on electron neutrinos only (and extrapolated to three flavors), 
ANITAII\cite{ANITAII} and a result from
the Auger experiment\cite{Auger}, based on tau neutrinos only (and extrapolated to three flavors). Models are shown as dashed or solid lines; experimental results as 
lines/points.}\label{fig:revised_UL}\end{figure}

Our flux limit is derived directly from the effective volume
$V_{eff}$, the livetime ${\cal L}$, and the event-finding
efficiency $\epsilon(\sim$0.64), 
which is the product of the
online software veto efficiency ($\epsilon_{online}\sim$0.91)
and the offline analysis efficiency
($\epsilon_{offline}\sim$0.742).
Our 95\% C.L. flux bounds  
are shown in 
Fig. \ref{fig:revised_UL}. Compared to our previous result,
we have slightly more than doubled our sensitivity. 
The dominating factor in our sensitivity 
gain is from extended livetime.

As can be immediately seen from inspection of Figure \ref{fig:revised_UL}, RICE is still well below the sensitivity required
to conclusively probe the ``cosmogenic'' neutrino flux, expected from interactions of the ultra-high energy cosmic
baryonic flux (protons, neutrons, or nuclei) with the cosmic microwave background (CMB). In brief, that flux is calculable
by bootstrapping from the ultra-high energy charged cosmic ray particle flux at Earth, assuming some source composition (at the
extremes, either proton or iron nuclei) for those measured charged cosmic rays, then integrating over redshift using some 
evolution model to obtain the anticipated rate in the current epoch. Using the parameters from the first such complete model\cite{ESS2001},
assuming an all-proton composition,
and integrating over the RICE sensitivity and livetime, 
we obtain an estimated number of 0.084 neutrino detections over 
the livetime quoted herein. Other recent estimates, which assume large admixtures of iron in the cosmic all-charged
spectrum, result in estimated rates an order of magnitude smaller (0.0063 events\cite{Kotera}).

\section{The Future of in-ice UHE Neutrino Detection at South Pole}
Clearly, larger effective volumes are needed to definitively confront the entire suite of
extant neutrino flux models. Synoptic strategies (ANITA, e.g.) afford sensitive volumes of $3\times 10^6$ ${\rm km}^3$, albeit viewed at typical distances of
order 100-400 km from the event vertex, resulting in high neutrino detection thresholds due to the 1/R signal strength losses. In the embedded signal detection scheme (RICE, e.g.), the neutrino interaction vertex is typically `close', but the sensitive volume limited by the radio frequency ice attenuation length and the $\sim$2 km-thickness of cold, high RF transparent ice, suggesting an ``ideal'' geometry of multiple `stations' of antennas deployed at shallow depths and separated by distances of order the radio attenuation length, each capable of independently imaging a neutrino interaction.

Central to the in-ice detection scheme are favorable radio frequency ice properties and transparency.
By now, several measurements have redundantly established bulk ice attenuation lengths of order 1--2 km in the
frequency range of interest. 
Within the last 2--3 years, as relative gains in neutrino sensitivity diminished, and in anticipation of a next-generation successor experiment, the RICE mission has begun to focus on precise characterization of asymmetries in ice properties, particularly effects of internal scattering layers and inherent asymmetries in the single-ice-crystal dielectric tensor. Both of these can be probed using bistatic radar echo sounding techniques. In this approach, a high-gain transmitter horn antenna is placed at one location on the snow surface, and the internal reflections from both within the snow, as well as the bedrock, are recorded by a second high-gain receiver horn antenna. Geometric asymmetries in the ice response can be studied by rotating the azimuthal plane of polarization of the horn antennas.

\begin{figure}[htpb]\centerline{\includegraphics[width=15cm]{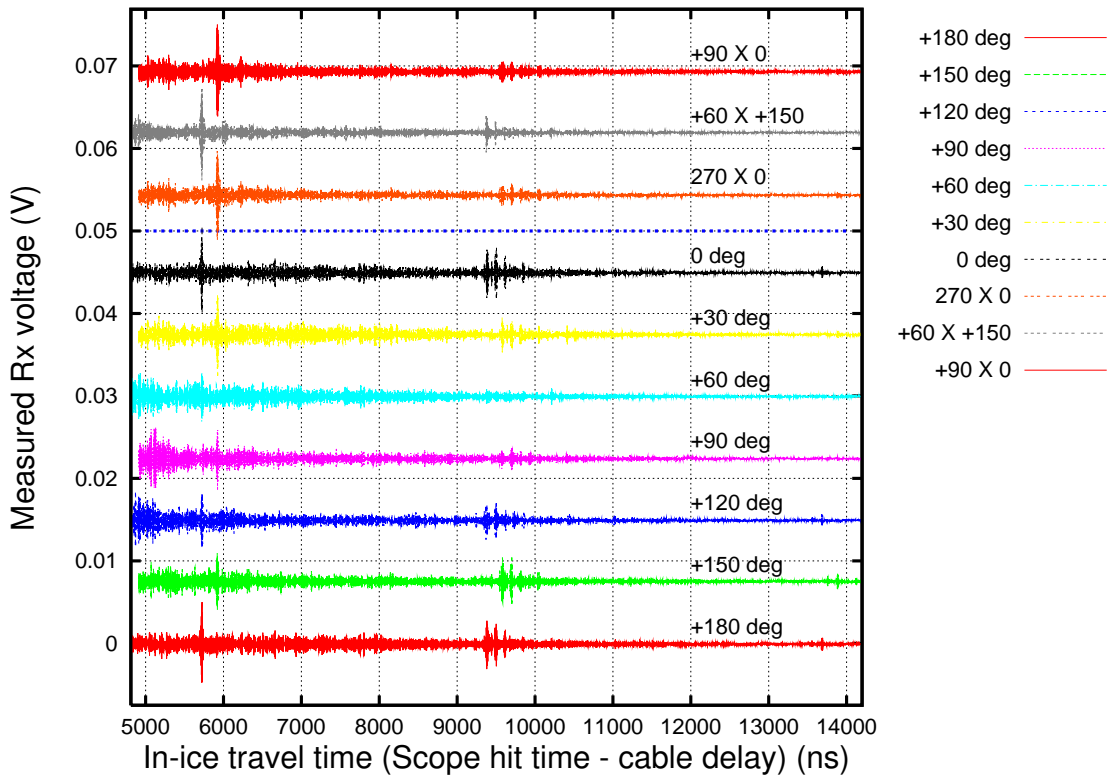}}
\caption{\it Ensemble of internal layer radar reflections observed, as a function of E-field polarization plane of vertically broadcast radio signals. In this Figure, 40000 waveform captures have been averaged; the coordinate system used is a local coordinate system for which the ice flow axis makes an angle of $153^\circ$ with respect to our zero degree convention. Azimuthal polarization angle is shown in the key; also included are `cross-polarized' (+60$\times$+150) orientation results, for which transmitter and receiver horn are orthogonal to each other. Echo time is shown horizontally, and approximately translates to depth via: depth [km]$\approx$ t[ns]/12000. For visual clarity, successive vertical traces have been offset by
$\pm$100 ns; reflection structure is actually synchronous to within 1 ns.}\label{fig:InternalLayers}\end{figure}

\begin{figure}[htpb]\centerline{\includegraphics[width=15cm]{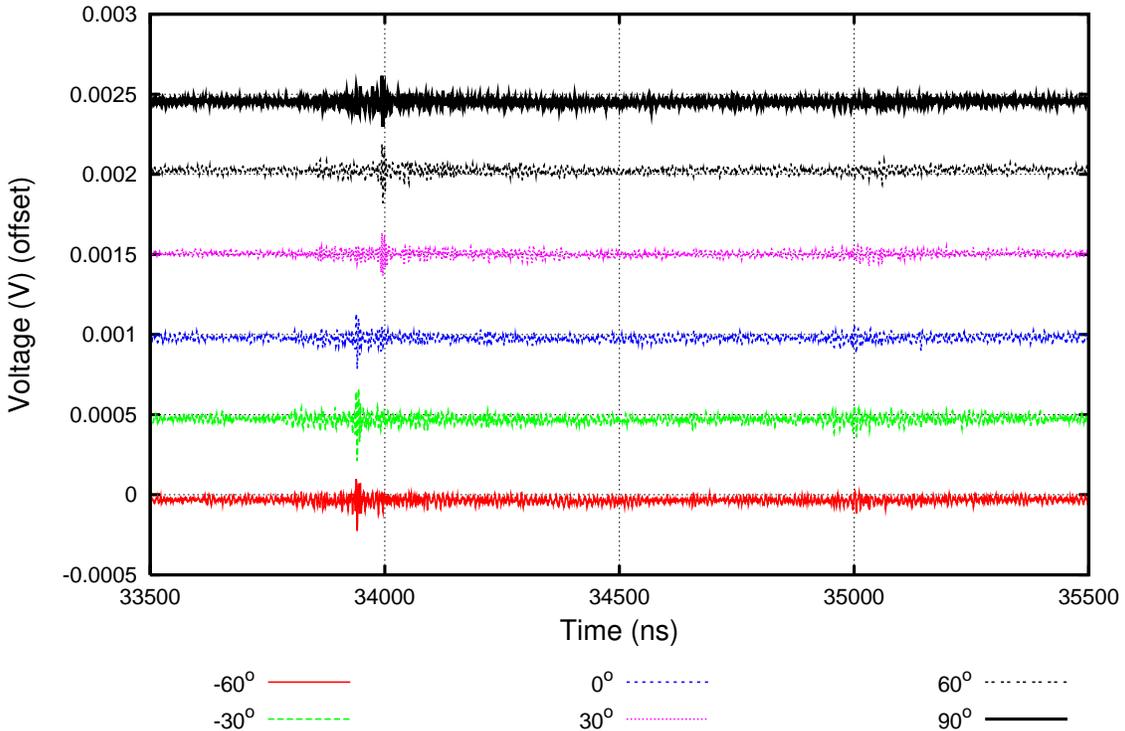}}\caption{\it Ensemble of bedrock radar reflections observed, as a function of E-field polarization plane of vertically broadcast radio signals. }\label{fig:BedRefl}\end{figure}

Figures \ref{fig:InternalLayers} and \ref{fig:BedRefl} show the measured reflections for times prior (Fig. \ref{fig:InternalLayers}) and corresponding (Fig. \ref{fig:BedRefl}) to the expected time for the bedrock echo, at a depth of 2850 m. Both show strong dependence of received signal with broadcast azimuthal angle, although only the latter shows the time delay between two polarizations indicative of a difference in index-of-refraction with orientation, i.e., birefringence. Quantitatively, the echo amplitudes observed from internal reflections are typically 20--40 dB reduced compared to those expected from a ``perfect mirror'', consistent with the characteristics of reflections from acid layers embedded within the ice itself. 
Frequency analysis of those reflections additionally corroborate the expected 1/f amplitude dependence of acid layer reflections.
Note that, for both these Figures, we have averaged over 10K--40K waveform captures to enhance the signal-to-noise ratio, corresponding to a reduction in the incoherent noise by (typically) at least two orders of magnitude. None of these reflections would therefore be visible in a single ``event'', such as an in-ice neutrino interaction.

Lab studies have shown that completely aligned ice crystals convey radio waves with approximately 1.7\% reduced speeds for propagation transverse to the plane containing that crystal (the $\hat{c}$-axis).
The time lag between the top three traces vs. the bottom three traces shown in Figure \ref{fig:BedRefl} corresponds to approximately 50 ns, over a total propagation time of 34000 ns, i.e., a birefringent asymmetry of order 0.15\% between wavespeed propagation along the ordinary (fast-) vs. extraordinary (slow-) axes. We can additionally use the {\it lack} of any asymmetry observed in Fig. \ref{fig:InternalLayers} to conclude that birefringence is a feature only of the lower (warmer) half of the ice sheet at South Pole, at a level of 0.25\% asymmetry. Taken together, these results indicate that losses due to internal radio layers will not result in appreciable loss of signal for neutrino-induced radio signals received in future experiments, and also that dimunition of peak signal strength due to birefringence will similarly be noticeable in only $\sim$5\% of all neutrino detection geometries.

\begin{figure}[htpb]\centerline{\includegraphics[width=15cm]{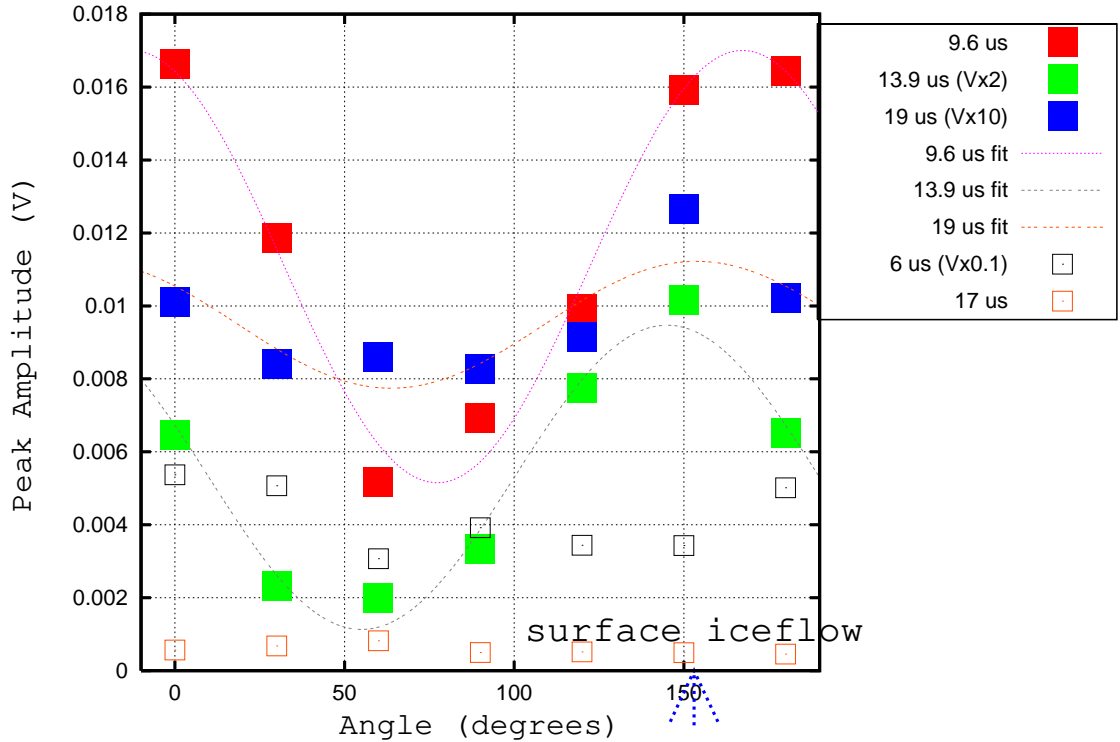}}\caption{\it Peak amplitude dependence of internal layer reflection, indexed by echo time, as a function of signal polarization. Note the correlation of phase with the direction of local ice flow direction, indicated by the dashed blue arrow.}\label{fig:A_v_phi}\end{figure}

Somewhat interestingly, Figure \ref{fig:InternalLayers} shows a marked dependence of peak measured reflected amplitude, as a function of azimuth.
Given that the only `preferred' horizontal direction is defined by the ice flow axis, it is natural to consider correlations between the amplitude variations observed in both the internal layer and also bedrock reflections, with the known bulk motion of the ice sheet. Fig. \ref{fig:A_v_phi} shows a strong correlation in three of the five most prominent observed internal layer echoes with the ice sheet flow direction and suggests that the internal acid layers are likely aligned (similar to a diffraction grating) by the local ice flow.

\subsection{Dependence of attenuation length with depth}
The observed echo amplitudes 
shown in Figure \ref{fig:InternalLayers}
are largely determined by three factors: the intrinsic reflectivity of each
layer, the diminution of signal power $P_{signal}$ with distance, and attenuation of the signal due to ice absorption.
For the directional horn antennas used in this experiment, $P_{signal}\propto r^{-\alpha}$, with 1$<\alpha<$2.
If we assume
approximately equivalent reflection coefficients for all observed
internal layers, we can determine a `local' amplitude attenuation length
between the first three, and last three layers, as shown in Table \ref{tab:localLatten},
by direct application of the Friis equation\citep{Friis}, and using the azimuth-averaged values of amplitude. For this calculation, we take $\alpha$=2; 
assuming a cylindrical-flux tube with no transverse spreading ($\alpha$=1) gives values 
approximately 20\% smaller than those presented in Table \ref{tab:localLatten}.
Our calculations are consistent with
the expected warming of the ice sheet from the bedrock below, 
and the corresponding reduction in attenuation length with
increasing temperature.

\begin{table}[htpb]
\begin{center}
\begin{tabular}{c|c|c|c} 
      & 13.9$\mu$s& 17.2$\mu$s& 19.6$\mu$s\\ \hline
6$\mu$s& 3348 m  & 1521 m   & 1514 m \\ 
9.6$\mu$s& 1170 m & 867 m & 964 m \\
13.9$\mu$s& & 643 m & 849 m \\ \hline
\end{tabular}
\end{center}
\caption{\it Inter-layer attenuation lengths, calculated from amplitudes
measured for returns, and assuming uniform reflectivity of all layers,
as discussed in text. Estimated systematic errors are of order 25--30\%.
First column indicates first reflecting layer; successive
columns indicate second reflecting layer used to calculate attenuation length
via Friis Equation. These results affirm the expectation that primary neutrino sensitivity is poorest
in the warm ice near the bedrock.}
\label{tab:localLatten}
\end{table}


\subsection{Future Plans for Radioglaciology}
Thus far, virtually all information on the radio frequency response of ice sheets has been derived using vertically broadcast signals. In the austral summer of 2011-12, RICE hardware will be used to broadcast RF at largely oblique angles, which will provide information on the ice response away from the ${\hat c}$-axis, and more typical of the geometry of the neutrino signals to be detected by in-ice experiments such as RICE. As illustrated above, all RICE studies done thus far, however, substantiate the basic premise of such detectors: that the excellent RF properties of cold polar ice imply that englacial neutrino detectors provide the most cost effective technique for confronting the cosmogenic neutrino flux.

\subsection{The ARA Experimental Initiative}
During the austral summer of
2010-11, initial deployments of the next generation of neutrino detection hardware, 
realized as
the recently funded Askaryan Radio Array (ARA\cite{ARA}), were
made at the South Pole. 
The first ARA deployment, in the form of a relatively shallow (20--30 m)
``testbed'' prototype has already demonstrated 20 arc-minute
angular reconstruction of calibration antennas, as well as sensitivity to 
variations in the received galactic noise and RF emissions from solar flares. An ambitious proposal,
including 14 institutions from eight countries has been submitted to develop
an autonomously-powered (and therefore, arbitrarily scaleable) experiment 
capable of initially defining the cosmogenic neutrino flux, and, over the timescale
of a decade, eventually performing statistical characterization of that flux.

Comprising 37 stations, each individually with
an energy reach approximately an order of magnitude below that
of  RICE, ARA will achieve a 
nearly 100-fold improvement in total effective volume,
achieved via:
\begin{enumerate}
\item direct digitization at the sensor (antenna) rather than on the surface, eliminating the 
$\sim$15 dB signal losses typically incurred by conveyance through coaxial cable.
\item order-of-magnitude
reduction of the geometric scale of each ``station'' from the 200-m typical of RICE such that
the temporal signal coincidence window can similarly be narrowed by a comparable factor of 10.
\item Siting of the experiment several km from the main South Pole station itself, resulting in
considerably lower ambient noise rates. The remote ARA deployment site exhibits virtually none of the anthropogenic, or wind-generated RFI that plaged the RICE data sample.
\item Extension of the lower-frequency limit of the antenna response from the current 250 MHz to 
$\sim$150 MHz, resulting in improved response to off-Cherenkov-peak signals.
\item ``Optimized'' antenna receiver placement, as opposed to the requirement that RICE co-deploy
in boreholes being drilled for the AMANDA experiment.
\end{enumerate}

The 2011-12 austral season will include the first deployment of a full-fledged ARA station (``ARA-1''); that single station will, in one year, have equivalent neutrino sensitivity to the ten years of RICE data accumulated thus far and reported herein. The first results on neutrino searches from the ARA testbed should be forthcoming within the next few months.

\section*{Acknowledgments}
The authors particular 
thank Chris Allen (U. of Kansas) 
for very helpful discussions, as well as our colleagues on the
RICE and ANITA experiments. We also thank Andy Bricker of
Lawrence High School (Lawrence, KS) for his assistance working
with the Lawrence and Free State High School
students. We also thank the winterovers at South Pole Station
(Xinhua Bai, The Most Rev. Allan Baker, Philip Braughton, Christina Hammock, Michael Offenbacher, Mark Noske, Nicolas Hart-Michel, Robert Fuhrman, Flint Hamblin, and Nick Strehl) whose efforts were essential
to the operation of this experiment. Sean Grullon contributed the 
ROOT software code used to produce the upper limit compilation presented in this document.
This work was supported by
the National Science Foundation's Office of Polar Programs
(grant OPP-0826747) and QuarkNet programs.


\begin{thebibliography}{...}
\bibitem{Askaryan}G. A. Askaryan, Zh. Eksp. Teor. Fiz. {\bf 41}, 616 (1961); Soviet Physics JTEP {\bf 14}, 441 (1962).
\bibitem{rice06}I. Kravchenko {\it et al} (RICE), Phys. Rev. {\bf D73}, 082002 (2006), astro-ph/0601148.
\bibitem{SALSA}P. Gorham {\it et al.}, Nucl. Instr. Meth. {\bf A490}, 476, (2002); R. Milincic {\it et al.}, arXiv:astro-ph/0503353.
\bibitem{RAMAND}A. Butkevich {\it et al.}, Fisika Elementarnik Chastitz \& Atomnovo Yadra {\bf 29}, 660 (1998); 
A. V. Butkevich, L. G. Dedenko, S. Kh. Karaevsky, A. A. Mironovich, A. L. Provorov, and I. M. Zheleznykh,  Physics of Particles and Nuclei {\bf 29}, 266 (1998) (translation).
\bibitem[Barwick,~S.~and~others, 2010]{ARIANNA}L. Gerhardt {\it et al.}, arXiv:1005.5193, (2010).
\bibitem[Gorham,~P.W.,~and others, 2009]{ANITAinstr}P.W. Gorham {\it et al.}, 
Astropart. Phys. {\bf 32}, 10, (2009).
\bibitem{FORTE}N. Lehtinen {\it et al.}, Phys. Rev. {\bf D69}, 013008 (2004).
\bibitem{RITA}I. Kravchenko {\it et al.}, Nucl. Inst. and Meth. in Physics Research {\bf A10.1016}/j.nima.2010.11.046 (2010).
\bibitem{GLUE}P. Gorham {\it et al.}, Phys. Rev. Lett. {\bf 93}, 041101 (2004).
\bibitem[Clancy James, 2010]{LUNASKA}C. W. James {\it et al.}, Phys. Rev. {\bf D81}, 042003 (2010).
\bibitem{LOPES}J. Hoerandel {\it et al.}, Nucl. Instr. Meth. {\bf A630}, 171 (2011).
\bibitem{CODALEMA}D. Ardouin {\it et al.}, Int. J. Mod. Phys. {\bf A20}, 6869 (2005).
\bibitem{AERA}P. Abreu {\it et al.}, accepted for publication in NIM {\bf A}, arXiv:1101.4473
\bibitem{ANITAcr}S. Hoover {\it et al.}, Phys. Rev. Lett. {\bf 105}, 151101 (2010).
\bibitem{BlackwellLovell_1941}P. M. S. Blackett and A. C. B. Lovell, Proc. Roy. Soc. {\bf A177} 183 (1941).
\bibitem{slac-testbeam}D. Saltzberg {\it et al.}, Phys. Rev. Lett. {\bf 86}, 2802 (2001).
\bibitem{slac-salt04}P. Gorham {\it et al.}, Phys. Rev. {\bf D72}, 023002 (2005).
\bibitem{ZHS}E. Zas, F. Halzen, and T. Stanev, Phys. Lett.
{\bf B257}, 432 (1991); E. Zas, F. Halzen, and T. Stanev, Phys. Rev.
{\bf D45}, 362 (1992).
\bibitem{Alvarez-papers}J. Alvarez-Muniz {\it et al.}, Phys. Rev. {\bf D68}, 043001 (2003); Phys. Rev. {\bf D62}, 063001 (2000); arXiv:astro-ph/0512337.
\bibitem{SoebPRD}S. Razzaque {\it et al.}, Phys. Rev. {\bf D65}, 103002 (2002).
\bibitem{addendum}S. Razzaque {\it et al.}, Phys. Rev. {\bf D69}, 047101 (2004).
\bibitem{RomanJohn} R.V. Buniy and J.P. Ralston, Phys. Rev. {\bf D65} 016003 (2002).
\bibitem{AndresJaime}J. Alvarez-Muniz, A. Romero-Wolf, and E. Zas, arXiv:1106.6283, submitted to Phys. Rev. {\bf D}.
\bibitem{Shahid-hadronic}S. Hussain and D. McKay, Phys. Rev. {\bf D70}, 103003 (2004).
\bibitem{LPM}L. D. Landau and I. J. Pomeranchuk, Dokl. Akad. Nauk. SSSR {\bf 92}, 535 (1953), 92, 735 (1953); A. B. Migdal, Phys. Rev. {\bf 103}, 1811 (1956); J. Alvarez-Mu\~niz, R.A. V\'azquez and E. Zas, Phys. Rev. {\bf D61}, 023001 (2000).
\bibitem{spencer04}S. Mandal, S. Klein and J.D. Jackson, Phys. Rev. {\bf D72}, 093003 (2005).
\bibitem{rice03a}I. Kravchenko {\it et al.}, Astropart. Phys. {\bf 19}, 15 (2003).
\bibitem{rice03b}I. Kravchenko {\it et al.}, Astropart. Phys. {\bf 20},
195 (2003).
\bibitem{RICEnZ}I. Kravchenko, D. Besson, and J. Meyers, J. Glac.
{\bf 50}, 171 (2004).
\bibitem{RICEfaraday} D.Besson, R.Keast, R.Velasco, Astropart. Phys. {\bf 31},
 5, (2009).
\bibitem{RICEbiref10}I. Kravchenko {\it et al.}, Astropart. Phys. {\bf 34}, 755, (2011).
\bibitem{grb06}D. Besson {\it et al.}, Astropart. Phys. {\bf 26}, 367, (2007).
\bibitem{daniel}D. P. Hogan {\it et al.}, Phys. Rev. {\bf D78}, 075031 (2008).
\bibitem{HuegeRefs} T. Huege \& H Falcke, Astropart. Phys. {\bf 24}, 116 (2005).
\bibitem{gazazian}E. Gazazian, K. Ispirian and A. Vardanyan, Proc. of 1st Int. Workshop on Radio Detection of High-Energy Particles, Ed. D. Saltzberg and P. Gorham. AIP Conf. Proc. {\bf 579} (2005).
\bibitem{VLF}\url{http://www-star.stanford.edu/~vlf/south_pole/south%20pole.htm}
\bibitem{Stecker2005}F. W. Stecker. Phys. Rev. {\bf D72}, 107301, (2005). 
\bibitem{Mucke2003}A. Mucke {\it et al.}, Astropart. Phys. {\bf 18}, 593, (2003). 
\bibitem{Soeb2003}S. Razzaque, P. M´eszaros, and E. Waxman. Phys. Rev. {\bf D68}, 083001, (2003). 
\bibitem{ESS2001}R. Engel, D. Seckel, and T. Stanev. Phys. Rev. {\bf D64}, 093010, (2001).
\bibitem{Ina2008}Enberg, Reno, and Sarcevic, Phys. Rev. {\bf D78}, 043005 (2008). 
\bibitem{Naumov2001}G. Fiorentini, V.A. Naumov, F.L. Villante Phys. Lett. {\bf B510}, (2001).
\bibitem{Bartol2004}T. Gaisser {\it et al.}, Phys. Rev. {\bf D70}, 023006, (2004).
\bibitem{Honda2006}Honda {\it et al.}, Phys. Rev. {\bf D75}, 043006 (2007).
\bibitem{Julia2005}J. Becker, Biermann, Rhode, Astropart. Phys. {\bf 23}, 355, (2005).
\bibitem{JohnKelly2009}R. Abassi {\it et al.}, Phys. Rev. {\bf D79}, 102005 (2009).
\bibitem{AMII09}R. Abbasi {\t et al.}, arXiv:1004.2357, accepted for publication in Astropart. Phys. 
\bibitem{AMII10}A. Achterberg {\it et al.}, Phys. Rev. {\bf D76}, 042008 (2007). 
\bibitem{AM08}M. Ackerman {\it et al.}, Astrophys. J. {\bf 675}, 1014 (2008).
\bibitem{HiRes}R. U. Abbasi {\it et al.}, Astrophys. J. {\bf 684}, 790 (2008).
\bibitem{ANITAII}P. W. Gorham {\it et al.}, Phys. Rev. {\bf D82}, 022004 (2010), and {\it erratum:}
P. W. Gorham {\it et al.}, \url{arXiv:1011.5004} (submitted to Phys. Rev. D).
\bibitem{Auger}J. Abraham {\it et al.}, Phys. Rev. {\bf D79}, 102001, (2009). 
\bibitem{Kotera}K. Kotera, D. Allard, A. V. Olinto, JCAP {\bf 1010}, 013 (2010).
\bibitem{Friis}http://en.wikipedia.org/wiki/Friis\_transmission\_equation
\bibitem{Falcke05} H. Falcke {\it et al.}, Nature {\bf 435}, 313 (2005).
\bibitem{ARA}P. Allison {\it et al.} (The ARA Collaboration), \url{arXiv:1105.2854}, submitted to Astropart. Phys.
\end{thebibliography}
\end{document}